\documentclass[twocolumn,prb]{revtex4}
\usepackage{graphicx}
\usepackage{dcolumn}
\usepackage{bm}
\usepackage{amsmath}
\usepackage{amssymb}

\begin{document}

\preprint{}
\title{Anomalous features of the proximity effect in triplet superconductors}
\author{Y. Tanaka$^{1,2}$, Y. Asano$^{3}$ A. A. Golubov$^{4}$ and S.
Kashiwaya$^{5}$}
\affiliation{$^1$Department of Applied Physics, Nagoya University, Nagoya, 464-8603,
Japan \\
$^2$ CREST, Japan Science and Technology Corporation (JST) Nagoya, 464-8603,
Japan \\
$^3$ Department of Applied Physics, Hokkaido University, Sapporo 060-8628,
Japan \\
$^4$Faculty of Science and Technology, University of Twente, 7500 AE,
Enschede, The Netherlands \\
$^5$NeRI of National Institute of Advanced Industrial Science and Technology
(AIST), Tsukuba, 305-8568, Japan }
\date{\today}

\begin{abstract}
Anomalous features of the proximity effect specific to diffusive normal
metal / triplet superconductor (DN/TS) junctions are studied within the
quasiclassical Green's function formalism. The pair amplitude $f_{N}$ as a
function of energy $\varepsilon$ in DN, satisfies an anomalous relation, $%
f_{N}(\varepsilon)=-f^{*}_{N}(-\varepsilon)$, contrary to that in singlet
superconductor junctions case where $f_{N}(\varepsilon)=f^{*}_{N}(-%
\varepsilon)$. Such an unusual $\varepsilon$ dependence is responsible for a
zero energy peak in the local density of states and is a source of an
anomalous penetration of the applied magnetic field into DN. %
Our results are relevant to Sr$_{2}$RuO$_{4}$, which is considered to have a
$p_{x}+ip_{y}$-wave symmetry.
\end{abstract}

\pacs{PACS numbers: 74.20.Rp, 74.50.+r, 74.70.Kn}
\maketitle



%

%




Nowadays spin-triplet superconductivity is a fascinating and an
important issue in condensed matter physics. 
The existence of triplet superconductivity has become promising after 
a series of experiments in Sr$ _{2}$RuO$_{4}$~\cite{Maeno,Maenonew}. 
Other possible triplet superconductors are UPt$_{3}$ \cite{tou} and
(TMTSF)$_{2}$X \cite{Lee}. Phase sensitive phenomena ~\cite{Tsuei}
in junctions involving triplet superconductor is an important and
relatively unexplored subject. 
One open question is how the proximity effect manifests itself in junctions 
between triplet superconductors and normal metals. \par
The proximity effect can be described in terms of the coherence between
electrons and Andreev reflected holes in a normal metal.
In unconventional superconductors, the \textit{internal phase} reflecting
sign change of a pair potential significantly affects charge transport since
this sign change is a source of a midgap Andreev resonant state (MARS) at
the surface of a superconductor. The origin of the MARS~\cite{Buch} is the
interference effect of a quasiparticle at the interface~\cite{Kashi00,TK95}.
In ballistic case, MARS leads to the formation of zero energy peak (ZEP) in
local density of state (LDOS). However, the interference is destructive for $%
d$-wave superconductor /diffusive normal metal (DN) junctions \cite{PRL2003}
and therefore no ZEP is expected in LDOS in DN. On the other hand, in
diffusive normal metal / triplet superconductor (DN/TS) junctions the MARS
penetrates into DN~\cite{Tanaka2004} and ZEP occurs in the LDOS in DN.

Recently, a search of new phenomena in triplet superconductors has attracted
a lot of attention. For example, recent experiments indicated possible
existence of MARS in the tunneling spectroscopy of Sr$_{2}$RuO$_{4}$ and Ru
micro inclusion system~\cite{Mao}. Therefore, it is important to explore the
physics of proximity effect in DN/TS junctions and to address various
situations relevant for the actual junctions. %

%
In the present paper, we study the anomalous proximity effect in N/DN/TS and
TS/DN/TS junctions (N denotes normal electrode) in the framework of the
Keldysh-Nambu quasiclassical Green's function formalism established in
previous work~\cite{PRL2003,Tanaka2004,Tanaka2005}. We find a crucial
relation in pair amplitudes in DN, $f_{N}(\varepsilon )=-f_{N}^{\ast
}(-\varepsilon )$ in contrast to a relation $f_{N}(\varepsilon )=f_{N}^{\ast
}(-\varepsilon )$, which holds in DN attached to a singlet superconductor.
Here $\varepsilon $ is the quasiparticle energy measured from the Fermi
level. The function $f_{N}(\varepsilon )$ is commonly used in the
quasiclassical Green function theory and characterizes the pair amplitudes
in DN due to the proximity effect \cite{Volkov}. The property $%
f_{N}(\varepsilon )=-f_{N}^{\ast }(-\varepsilon )$ is responsible for the
penetration of MARS into N/DN/TS junctions which provides the ZEP in LDOS.
Moreover, we predict an unusual response of DN to an external magnetic
field. We show that the magnetic field spatially oscillates in DN instead of
usual Meissner screening. These unusual effects do not exist in the
conventional proximity effect in singlet superconductors. We also show that
in TS/DN/TS junctions, the proximity effect is suppressed when the external
phase difference across the junction $\varphi $ is $0$ due to the
destructive interference of MARS's from the two superconductors.
On the other hand, at $\varphi =\pi $ the constructive interference bridges
the two MARS's in DN.

%
%
The model of the system is shown in Fig. 1. We consider a junction
consisting of a normal metal and a superconducting reservoir connected by a
quasi-one-dimensional diffusive conductor (DN) with a length $L$ much larger
than the mean free path. The interface between DN and TS has a resistance $%
R_{b}$ while the DN/N interface has a resistance $R_{b^{\prime }}$. We
assume flat interfaces and the insulating barriers at the two interfaces are
modeled by the delta-function~\cite{Tanaka2005}. We start from the case
where the pair potential is chosen to be $p_{x}+ip_{y}$-wave pairing
symmetry \cite{Ueda} which is the promising candidate of superconductivity
in Sr$_{2}$RuO$_{4}$ \cite{Maeno}. As regards the spin structure of the Cooper pair, we choose $S_{z}=0$ . The positions of the DN/N interface and the DN/TS
interface are denoted as $x=0$ and $x=L$, respectively. We calculate the
Keldysh-Nambu quasiclassical Green's functions by using the general boundary
conditions given in Eqs.~(4) and (5) in Ref.~\onlinecite{Tanaka2005}. To
calculate the LDOS in DN, we focus on the retarded part of the Nambu-Keldysh
Green's functions. For $p_{x}+ip_{y}$-wave pairing symmetry, the
quasiclassical Green's function in TS is given by
\begin{equation}
\hat{R}_{2\pm }=f_{1\pm }\hat{\tau}_{1}+f_{2\pm }\hat{\tau}_{2}+g\hat{\tau}%
_{3}
\end{equation}%
with $f_{1\pm }=\Delta _{1,\pm }/\sqrt{\Delta _{0}^{2}-\varepsilon ^{2}}$, $%
f_{2\pm }=-\Delta _{2,\pm }/\sqrt{\Delta _{0}^{2}-\varepsilon ^{2}}$, $%
g=\varepsilon /\sqrt{\varepsilon ^{2}-\Delta _{0}^{2}}$, where $\Delta
_{1\pm }$ and $\Delta _{2\pm }$ are given by $\Delta _{1\pm }=\pm \Delta
_{0}\cos \phi $ and $\Delta _{2\pm }=\Delta _{0}\sin \phi $, respectively.
Here $\Delta _{1,+}+i\Delta _{2,+}$ and $\Delta _{1,-}+i\Delta _{2,-}$ are
the effective pair potentials for a quasiparticle with an injection angle $%
\phi $ and $(\pi -\phi )$, respectively. The Green's function in DN is
expressed by
\begin{equation}
\hat{R}_{N}(x)=\sin \theta \hat{\tau}_{2}+\cos \theta \hat{\tau}_{3}
\label{Usadel}
\end{equation}%
with the proximity parameter $\theta$. 
Since the $\hat{\tau}_{3}$
component of the matrix current in Eq. (4) of
Ref. (\onlinecite{Tanaka2005}) vanishes due to the absence of the
Josephson current after the angular average of the matrix current as
a function of $\varphi $, the $\hat{\tau}_{1}$ component
of the Green's function in DN is vanishing. As discussed in 
Ref.~(\onlinecite{Tanaka2005}), function $\theta $ in DN is determined by the
Usadel equation
\begin{equation}
D\frac{\partial ^{2}}{\partial x^{2}}\theta +2i\varepsilon \sin \theta =0
\label{Eq_theta}
\end{equation}%
supplemented by the boundary conditions \cite{KL,PRL2003,Nazarov2}
\begin{equation}
\frac{LR_{b}}{R_{d}}\left. \left( \frac{\partial \theta }{\partial x}\right)
\right\vert _{x=L_{-}}=\langle F_{1}\rangle ,\;\;\frac{LR_{b^{\prime }}}{%
R_{d}}\left. \left( \frac{\partial \theta }{\partial x}\right) \right\vert
_{x=0_{+}}=\langle F_{2}\rangle ,  \label{bc}
\end{equation}%
%
%
where%
\begin{align}
F_{1}=& \frac{4T_{1}[(1+T_{1}^{2})(A-Bf_{1}i)+2T_{1}(AB-if_{1})]}{%
[(1+T_{1}^{2})+2T_{1}B]^{2}-(1-T_{1}^{2})^{2}f_{1}^{2}},  \label{f1} \\
F_{2}=& \frac{2T^{\prime }\sin \theta _{0}}{2-T^{\prime }+T^{\prime }\cos
\theta _{0}},  \label{f2}
\end{align}%
with $A=(\cos \theta _{L}f_{2}-\sin \theta _{L}g)$, $B=(\sin \theta
_{L}f_{2}+\cos \theta _{L}g)$, $T_{1}=T_{\phi }/(2-T_{\phi }+2\sqrt{%
1-T_{\phi }})$ and $T^{\prime }=T_{\phi}^{\prime }$.

Here $D$ is the diffusion constant i DN and $\langle \ldots \rangle $
denotes the average over angle $\phi $ defined as follows
\begin{equation}
<F_{1(2)}(\phi )>=\int_{-\pi /2}^{\pi /2}d\phi \cos \phi F_{1(2)}(\phi
)/\int_{-\pi /2}^{\pi /2}d\phi T_{\phi }\cos \phi ,  \label{average}
\end{equation}%
$R_{d}$ is the resistance of DN, $\theta _{L}$ and $\theta _{0}$ are the
values of $\theta $ at $x=L$ and $x=0$, respectively. In the above, $%
f_{1}=f_{1+}=-f_{1-}$ and $f_{2}=f_{2+}=f_{2-}$ are satisfied. The
transparencies at the DN/TS and N/DN interface are given by
\begin{equation}
T_{\phi }=\frac{4\cos ^{2}\phi }{Z^{2}+4\cos ^{2}\phi },\;\;T_{\phi
}^{\prime }=\frac{4\cos ^{2}\phi }{Z^{\prime }{}^{2}+4\cos ^{2}\phi },
\end{equation}%
where $Z$ and $Z^{\prime }$ denote the barrier parameter at the interfaces.

At first we discuss the unusual $\varepsilon $ dependence of $\theta $.
%
It is easy to confirm relations $f_{1(2)}(-\varepsilon ,\phi
)=f_{1(2)}^{\ast }(\varepsilon ,\phi )$, $g(-\varepsilon ,\phi )=g^{\ast
}(\varepsilon ,\phi )$, $f_{1}(\varepsilon ,-\phi )=f_{1}(\varepsilon ,-\phi
)$, $f_{2}(\varepsilon ,-\phi )=-f_{2}(\varepsilon ,-\phi )$, and $%
g(\varepsilon ,-\phi )=g(\varepsilon ,\phi )$ by the definition of $f_{1,2}$
and $g$. By applying these relations to Eqs.~(\ref{bc})-(\ref{f2}), we
immediately find
\begin{equation}
\theta (-\varepsilon )=-\theta ^{\ast }(\varepsilon ).  \label{fep}
\end{equation}%
It can be shown from Eq. (22) in Ref.~(\onlinecite{Tanaka2005}) that this
equation holds for triplet junctions when time reversal symmetry is
preserved ~\cite{Tanaka2005}. We note that $\theta (-\varepsilon )=\theta
^{\ast }(\varepsilon )$~\cite{Volkov} is satisfied in any singlet junction
which preserves the time reversal symmetry. %
Thus, Eq.~(\ref{fep}) is an important and a unique relation in triplet
junctions. This relation leads to unusual properties of the proximity effect
in triplet junctions. In what follows, we explicitly describe $x$ in $\rho
(\varepsilon ,x)$ and $f_{N}(\varepsilon ,x)=\sin \theta (\varepsilon ,x)$.
\begin{figure}[bh]
\begin{center}
\scalebox{0.7}{
\includegraphics[width=8.0cm,clip]{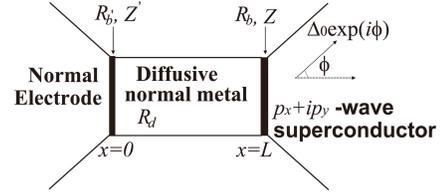}
}
\end{center}
\caption{ Schematic illustration of the model of the N/DN/TS junction with $%
p_{x}+ip_{y}$-wave pairing }
\label{fig:1}
\end{figure}
In Fig.~2, the local density of states $\rho (\varepsilon ,x)=\mathrm{Real}%
[\cos \theta (\varepsilon ,x)]$ and $f_{N}(\varepsilon ,x)$ with $\theta
(\varepsilon ,x)=\theta (\varepsilon )=\theta $ at the center of DN (i.e., $%
x=L/2$ ) in N/DN/TS junctions are plotted for $p_{x}$ and $p_{x}+ip_{y}$,
where $Z=1$, $Z^{\prime }=1$, and we fix a ratio $R_{d}/R_{b}$ at 1. The
interface resistance at N/DN ($R_{b^{\prime }}$) is a parameters of the
calculation. In order to achieve the convergence of LDOS in numerical
calculation, we replace $\varepsilon $ with $\varepsilon +i\gamma $ where $%
\gamma $ is small parameter, which we choose to be $0.005\Delta _{0}$. As
shown in Fig.~2(a), the LDOS has a ZEP, which indicates that MARS due to the
proximity effect exists deep in the DN. The height of the zero energy peak
(ZEP) of LDOS increases with the decrease of $R_{d}/R_{b^{\prime }}$.
%
This is because the pair amplitude $f_{N}(\varepsilon ,x)$ is more strongly
confined within DN for larger $R_{b^{\prime }}$. In $p_{x}$-wave pairing
symmetry, the same tendency is found in $\rho (\varepsilon ,L/2)$ as shown
in Fig.~2(b). In $s$-wave junctions, in contrast to the triplet case, the
amplitude of LDOS around $\varepsilon =0$ decreases with the decrease of $%
R_{d}/R_{b^{\prime }}$ as shown in (c). For small $R_{d}/R_{b^{\prime }}$,
quasiparticles in N are separated from those in DN and the minigap exists in
DN. %
%
With the increase of $R_{d}/R_{b^{\prime }}$ the states within the minigap
are filled in and the LDOS around $\varepsilon =0$ increases.

It is possible to understand the unusual $\varepsilon $ dependence of LDOS
for N/DN/TS junctions with $p_{x}$-wave symmetry in the limit $%
R_{d}/R_{b^{\prime }}\rightarrow 0$. We concentrate on the case with $%
0<\varepsilon <<\Delta _{0}$ and $T_{\phi }<<1$. The spatial dependence of $%
\theta $ has a form $\theta =\theta _{L0}+(x-L)\theta _{1}/L+(x-L)^{2}\theta
_{2}/(2L^{2})$. Substituting $f_{1}=\Delta _{0}\cos \phi /\sqrt{\Delta
_{0}^{2}\cos ^{2}\phi -\bar{\varepsilon}^{2}}$, $f_{2}=0$, and $%
g=\bar{\varepsilon}/\sqrt{\bar{\varepsilon}^{2}-\Delta _{0}^{2}\cos ^{2}\phi }$
with $\bar{\varepsilon}=\varepsilon +i\gamma $ into Eq. (\ref{f1}), we find $%
<F_{1}>\sim \Delta _{0}(\sin \theta _{L0}+i\cos \theta _{L0})/\gamma $.
Further, using Eqs. (\ref{Eq_theta}) and (\ref{bc}), we find $\cos \theta
_{L0}$ and $\rho (\varepsilon,x)$
\begin{equation}
\cos \theta _{L0}\sim \sqrt{\frac{\Delta _{0}R_{d}D}{4iL^{2}R_{b}\gamma
\varepsilon }},\ \ \rho (\varepsilon ,x)\sim \sqrt{\frac{\Delta _{0}R_{d}D}{%
8L^{2}R_{b}\gamma \varepsilon }}
\end{equation}%
It is remarkable that the $\rho (\varepsilon ,x)$ is proportional to $\sqrt{%
1/(\varepsilon \gamma )}$. This unusual $\varepsilon $ dependence is
specific to DN/TS junctions.

\begin{figure}[bh]
\begin{center}
\scalebox{0.4}{
\includegraphics[width=20.0cm,clip]{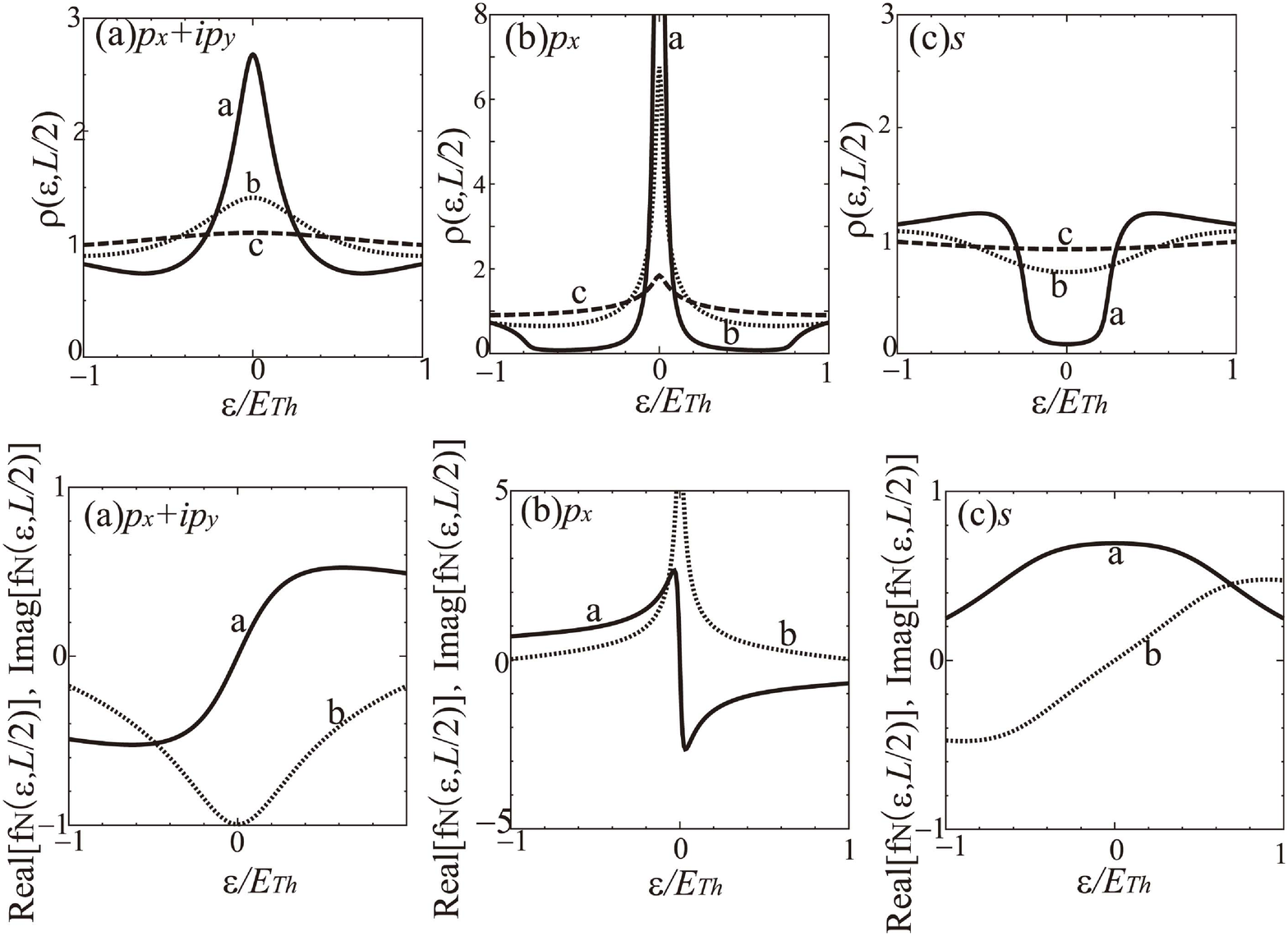}
}
\end{center}
\caption{ In the upper panel, local density of states $\protect\rho (\protect%
\varepsilon ,L/2)$ is plotted for N/DN/S junctions for various
superconducting state with $Z=1$, $R_{d}/R_{b}=1$, $Z^{\prime }=1$ and $%
E_{Th}=0.25\Delta _{0}$, respectively. (a)$p_{x}+ip_{y}$-wave, (b)$p_{x}$%
-wave and (c)$s$-wave pairing. a, $R_{d}/R_{b^{\prime }}=0.01$; b, $%
R_{d}/R_{b^{\prime }}=1$; and c, $R_{d}/R_{b^{\prime }}=100$. The corresponding pair
amplitude $f_{N}(\protect\varepsilon ,L/2)$ with $R_{d}/R_{b^{\prime }}=1$
are plotted for (a)$p_{x}+ip_{y}$-wave, (b)$p_{x}$-wave and (c)$s$-wave
pairing, respectively. a: $\mathrm{Real}[f_{N}(\varepsilon
,L/2)]$, b: $\mathrm{Imag}[f_{N}(\varepsilon ,L/2)]$. }
\label{fig:2}
\end{figure}
In the lower panels of Fig.~2, the $\varepsilon $ dependence of the pair
amplitude $f_{N}(\varepsilon ,L/2)$ is plotted for $R_{d}/R_{b}=1$. For
triplet junctions in (a) and (b), the real and imaginary parts of $%
f_{N}(\varepsilon ,L/2)$ are odd and even functions of $\varepsilon $,
respectively. On the other hand, in $s$-wave junctions, the real part of $%
f_{N}(\varepsilon ,L/2)$ is even function of $\varepsilon $ and the
imaginary part is odd function of $\varepsilon $. These features are
naturally understood from the fact that $\mathrm{Real}[f_{N}(\varepsilon
,L/2)]$ and $\mathrm{Imag}[f_{N}(\varepsilon ,L/2)]$ are respectively given
by $\sin \theta _{r}\cosh \theta _{i}$ and $\cos \theta _{r}\sinh \theta
_{i} $, where $\theta _{r}$ ($\theta _{i})$ is the real (imaginary) part of $%
\theta (\varepsilon ,L/2)$.

%
%
%
The unusual $\varepsilon $ dependence of $f_{N}(\varepsilon ,x)$ in DN/TS
junctions is a source of an anomalous response of DN to external magnetic
field. When a magnetic field is applied parallel to the interface, the
screening current $j(x)$ is given by\cite{Belzig}
\begin{equation}
j(x)=\pi e^{2}N(0)DT\sum_{\omega _{n}}\mathrm{Trace}[\hat{\tau}_{3}\hat{R}%
_{N}(x)[\hat{\tau}_{3},\hat{R}_{N}(x)]]A(x)
\end{equation}%
where $A(x)$ is the vector potential, $N(0)$ is the density of states in the
normal state at the Fermi level in DN, $\omega _{n}=\pi T(2n+1)$ and $T$ is
the temperature. As a result, the magnetic field in the DN behaves as $%
H(x)\sim \exp (-x/\lambda (x))$ with the local penetration depth $\lambda
(x) $, which is given by
\begin{equation}
\frac{1}{\lambda ^{2}(x)}=\frac{T\sum_{\omega_{n}}\sin ^{2}\theta (\omega _{n})}{%
\lambda _{0}^{2}},
\end{equation}%
as shown in Eq. (18) of Ref. (\onlinecite{Belzig}) or Eq. (2.5) of Ref. (%
\onlinecite{Narikiyo}) with $\lambda _{0}^{-2}=32\pi ^{2}e^{2}N(0)DT_{C}$,
where $T_{C}$ is the transition temperature of TS. In Fig.~3, the averaged
value of $\lambda ^{2}$, ($\bar{\lambda}_{av}^{2}=L/\int_{0}^{L}\frac{dx}{%
\lambda ^{2}(x)}$) is plotted as a function of temperatures ($T$), where $%
Z=1 $, $R_{d}/R_{b}=1$, $Z^{\prime }=1$, $R_{d}/R_{b^{\prime }}=1$ and $%
E_{Th}=0.25\Delta _{0}$. As is shown in Fig. 3 by curves c and d, $\bar{%
\lambda}_{av}^{2}>0$ for singlet junctions. Thus $\bar{\lambda}_{av}$ is a
real number and a magnetic field is screened by the usual Meissner effect in
the DN. On the other hand, in triplet junctions we find $\bar{\lambda}%
_{av}^{2}<0$ as shown by curves a and b. Therefore, the $\bar{\lambda}_{av}$
becomes a purely imaginary number for $p_{x}+ip_{y}$- and $p_{x}$-wave
junctions. Formally, this is the consequence of the fact that the pair
amplitude $f_{N}(i\omega _{N},x)$ is purely imaginary. It is a novel feature
of the anomalous proximity effect specific to triplet junctions that the
applied magnetic field is not screened in the DN region and is screened only
by the TS part.

\begin{figure}[bh]
\begin{center}
\scalebox{0.4}{
\includegraphics[width=10.0cm,clip]{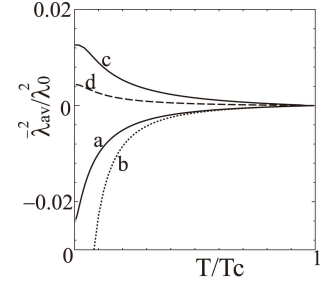}
}
\end{center}
\caption{ Averaged value of the Meissner screening length 
$\bar{\protect\lambda}_{av}^{2}$ is plotted for various superconducting state with $Z=1$, $%
R_{d}/R_{b}=1$, $Z^{\prime }=1$, $R_{d}/R_{b^{\prime }}=1$ and $%
E_{Th}=0.25\Delta _{0}$, respectively. (a)$p_{x}+ip_{y}$-wave, (b)$p_{x}$%
-wave, (c)$s$-wave and (d)$d_{x^{2}-y^{2}}+id_{xy}$-wave pairing. }
\label{fig:4}
\end{figure}
Let us now discuss the proximity effect in TS/DN/TS junctions by replacing
normal electrode (see Fig. 1) by triplet superconductor (TS). We study the
effect of an external phase difference $\varphi $ of the $p_{x}+ip_{y}$-wave
pair potentials on both sides of the junction on the LDOS in the DN layer.
We assume that the barriers at $x=0$ and $x=-L$ are the symmetric (i.e., $%
R_{b^{\prime }}=R_{b}$ and $Z^{\prime }=Z$). We consider the two situations,
$\varphi =0$ and $\varphi =\pi $. The boundary condition at $x=0$ is
\begin{equation}
\frac{LR_{b^{\prime }}}{R_{d}}\left. \left( \frac{\partial \theta }{\partial
x}\right) \right\vert _{x=0_{+}}=\langle F_{2}^{\prime }\rangle  \label{bc1}
\end{equation}%
where $F_{2}^{\prime }$ is given by
\begin{equation}
F_{2}^{\prime }=-\frac{4T_{1}[(1+T_{1}^{2})(A^{\prime }-B^{\prime
}f_{1}^{\prime }i)+2T_{1}(A^{\prime }B^{\prime }-if_{1}^{\prime })]}{%
[(1+T_{1}^{2})+2T_{1}B^{\prime }]^{2}-(1-T_{1}^{2})f_{1}^{\prime }{}^{2}}
\label{rbc}
\end{equation}%
with $A^{\prime }=(\cos \theta _{0}f_{2}^{\prime }-\sin \theta _{0}g)$ and $%
B^{\prime }=(\sin \theta _{0}f_{2}^{\prime }+\cos \theta _{0}g)$. For $%
\varphi =0$, we find $f_{1}^{\prime }=-f_{1}$, $f_{2}^{\prime }=f_{2}$ and
for $\varphi =\pi $, $f_{1}^{\prime }=f_{1}$, $f_{2}^{\prime }=-f_{2}$, with
$f_{1}=\Delta _{0}\cos \phi /\sqrt{\Delta _{0}^{2}-\varepsilon ^{2}}$, and $%
f_{2}=\Delta _{0}\sin \phi /\sqrt{\Delta _{0}^{2}-\varepsilon ^{2}}$. %
As in the case of N/DN/TS junctions, 
to achieve the convergence of LDOS in numerical
calculation, we replace $\varepsilon $ with $\varepsilon +i\gamma $ 
with small parameter $\gamma=0.005\Delta _{0}$.

Here, we concentrate on the LDOS at the center of DN, $\rho (\varepsilon
,L/2)$, plotted in Fig.~4 for $\varphi =0$ and $\varphi =\pi $. %
As shown in Fig. 4(a), for $\varphi =0$, $\rho (\varepsilon ,L/2)=1$ in the
whole $\varepsilon $ range. On the other hand, for $\varphi =\pi $, $\rho
(\varepsilon ,L/2)$ has a clear ZEP. To understand this difference, we look
at the property of $\theta $. By comparing the left and right boundary
conditions using Eq. (\ref{bc}) and (\ref{rbc}), we can show for $\varphi =0$
that $\theta _{L}=-\theta _{0}$ and $\frac{d\theta }{dx}\mid _{x=L_{-}}=%
\frac{d\theta }{dx}\mid _{x=0_{+}}$. Then, the resulting pair amplitudes
penetrating from the left side superconductor and the right side interfere
destructively. Triplet pair potentials with $p_{x}+ip_{y}$-wave symmetry
change sign under spatial inversion. The phase difference of $\pi $ between
the two pair amplitudes with MARS's penetrating from the two superconductors
causes a destructive interference in DN in the absence of an external phase
difference. On the other hand, for $\varphi =\pi $, the resulting $\theta $
satisfies $\theta _{L}=\theta _{0}$ and $\frac{d\theta }{dx}\mid _{x=L_{-}}=-%
\frac{d\theta }{dx}\mid _{x=0_{+}}$, and the interference becomes
constructive in DN. The corresponding plot is also shown in the $s$-wave
case. In this case, the constructive interference is enhanced for $\varphi
=0 $ and the resulting $\rho (\varepsilon ,L/2)$ has a minigap. On the other
hand, for $\varphi =\pi $, due to the destructive interference, the
resulting $\rho (\varepsilon ,L/2)$ is unity due to the absence of the
proximity effect at $x=L/2$.

\begin{figure}[bh]
\begin{center}
\scalebox{0.4}{
\includegraphics[width=16.0cm,clip]{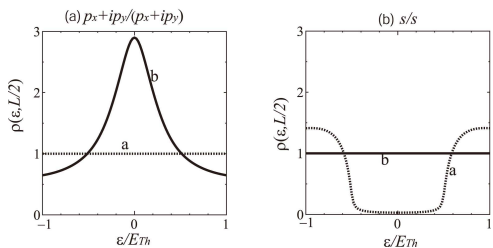}
}
\end{center}
\caption{ Spatial dependence of $\protect\rho(\protect\varepsilon,L/2)$ is
plotted for (a)$p_{x}+ip_{y}/p_{x}+ip_{y}$ junctions and (b)$s/s$ junctions
with $Z=Z^{\prime}=1$, $R_{d}/R_{b}=R_{d}/R_{b^{\prime}}=1$, and $%
E_{Th}=0.25\Delta_{0}$, respectively for a: $\protect\varphi=0$ and b: $%
\protect\varphi=\protect\pi$. }
\label{fig:5}
\end{figure}

In this paper we have studied anomalous features of the proximity effect in
triplet superconductor junctions in the framework of the quasiclassical
Green's function formalism. We have shown that the pair amplitudes $%
f_{N}(\varepsilon )$ obey the unusual relation (i.e., $f_{N}(\varepsilon
)=-f_{N}^{\ast }(-\varepsilon )$ in Eq.(\ref{fep})) in contrast to that in
singlet superconductor junctions. This unusual energy dependence induces the
ZEP of LDOS in DN and is also responsible for the unusual behavior of a
magnetic field in DN. The magnetic field is not screened in triplet
junctions and oscillates spatially in DN. In N/DN/TS junctions, the
magnitude of the ZEP increases with the increase of the interface resistance
between N and DN which is a result of the confinement of the MARS in the DN.
Therefore, to observe the predicted ZEP in LDOS in experiments on Sr$_{2}$RuO%
$_{4}$ and Ru micro inclusion systems 
a large resistance between N and DN is necessary. In TS/DN/TS junctions, the
proximity effect is suppressed when the external phase difference across
junctions $\varphi =0$ because two MARS's penetrating from the two
superconductors cancel each other. On the other hand, for $\varphi =\pi $,
the constructive interference bridges the two MARS's in DN. These results
imply the phase-sensitivity of MARS. Experimental observation of the
predicted unusual proximity effect in triplet superconductor proximity
systems would be of high interest.

%

%


\end{document}